# Green's functions of non-Hermitian systems at exceptional deficiency

Congwei Lu[1], Guancong Ma[1,2,†]

[1]Department of Physics, Hong Kong Baptist University, Kowloon Tong, Hong Kong, China

[2]Shenzhen Institute for Research and Continuing Education, Hong Kong Baptist University, Shenzhen 518000, China

[†]Email: phgcma@hkbu.edu.hk

**Abstract**

Exceptional deficiency is a newly discovered broadband non-Hermitian condition, at which the system's spectrum is entirely composed of exceptional points and all eigenvectors pairwise coalesce. Here, we analyze the steady-state and time-domain responses of non-Hermitian lattices at exceptional deficiency by considering their frequency and time-domain Green's functions. We show that the frequency-domain Green's functions are characterized by the emergence of second-order poles across the entire spectrum, producing broadband second-order super-Lorentzian line shapes, enhanced magnitude, and sequential $2\pi$ phase shifts across each resonance. These second-order poles also underpin unconventional responses to global dissipation, by which we uncover a loss-revival of defective skin effect: the skin modes that are "missing" due to the system's defectiveness at exceptional deficiency can re-emerge in the responses. In the time-domain responses, a linear-in-time amplification factor is identified in the evolution kernel, which can enhance skin-effect dynamics that would otherwise be suppressed by global loss. The amplification strength scales with the degree of defectiveness of the system. Our work establishes a theoretical framework for analyzing steady-state and dynamic signatures of exceptional deficiency, and opens new routes for dissipation-controlled broadband exceptional responses.

## 1 | Introduction

The non-Hermitian formalism has emerged as a powerful method for studying open systems whose properties hinge on non-conservative energy exchange [1, 2]. A peculiar property of non-Hermitian systems is that the eigenvalues, or the spectra, are generically complex. This leads to the non-Hermitian degeneracy known as exceptional points (EPs), characterized by the degeneracy of eigenvalues and the alignment of eigenvectors of individual states [2, 3]. Research on EPs has been thriving, fueled by fundamental intrigue in non-Hermitian topology [3, 4] and by promises in novel applications [5-7], such as lasing [8-11], high-performance sensors [12-20], robust long-distanced energy transfer [21, 22], chiral wave transfer [23-25], and invisibility [26, 27]. Importantly, many EP-induced phenomena and applications are underpinned by the system's response near or at the EP, which is described by the Green's function (GF). Poles of GF determine the line shapes of resonant peaks and the phase evolution in the spectrum, the latter is closely related to zeros of GF through wave interference. The poles and zeros together dominate the system's spectral responses. Previous theoretical investigations revealed that coalescence of eigenvectors produces a Jordan-block contribution to the GF at EP, which can contain higher-order poles in the GF. Such higher-order poles lead to super-Lorentzian response lines [28, 29]. These features can substantially alter the response properties, making the GF response an experimentally relevant signature of exceptional physics.

Despite the intriguing properties and advanced promises, EPs being degenerate points fundamentally limit their functionality to a narrow bandwidth and specific parameter regimes that often require fine-tuning to access. Recently, by considering a high-dimensional generalization of EP, a new non-Hermitian critical state called exceptional deficiency (ED) has been identified [30]. It is highlighted by the coalescence of two arbitrary-dimensional eigenspaces and the coincidence of their corresponding spectra. One unique characteristic of ED is that it allows EPs to form a spectral continuum that can potentially cover a large bandwidth. In this work, we systematically consider the steady-state and dynamical responses of non-Hermitian systems with semi-continuum spectra under the ED condition by studying the frequency- and time-domain GFs. We reveal several exotic features in the GF under the ED condition (EDGF). In the frequency-domain EDGF and under the proper form of excitation, the response is determined by second-order poles in the EDGF across the entire spectrum. By the interplay of ED-induced extended states and global loss, we discovered a new phenomenon called loss-revival of defective skin effect (LRDSE), in which the skin modes that are "missing" due to the system's defectiveness under the ED condition can re-emerge in the responses. In the time-domain EDGF, we identified a unique linear time-dependent amplification for all states. Both the LRDSE and the time-linear amplification uniquely arise from the condition of ED and have no counterpart in Hermitian or conventional non-Hermitian systems. Our work lays a foundation for future research exploring the properties of ED for new phenomena and applications.

**2 | Exceptional deficiency**

We first briefly lay down the concept of ED, which was first introduced in ref.[30]. Consider a generic model consisting of two subsystems, with the Hamiltonian given by

$$\mathcal{H} = \begin{pmatrix} H_A & \kappa \\ 0 & H_B \end{pmatrix}, \tag{1}$$

where $H_A$ and $H_B$ are each an $N$-dim diagonalizable Hamiltonian of two subsystems, and $\kappa$ represents the one-way hopping from $H_B$ to $H_A$. In previous investigations, it was shown that Hamiltonians of such a block-triangular form exhibited an anomalous non-Hermitian skin effect (NHSE) [31, 32], and were route towards higher-order EPs [28, 33]. But some additional conditions must be met for the ED to emerge. The left and right eigenvectors of $H_A$ and $H_B$ satisfy

$$H_A|\alpha_m\rangle = E_A^m|\alpha_m\rangle, \langle\langle\alpha_m|H_A = \langle\langle\alpha_m|E_A^m,$$

$$H_B|\beta_m\rangle = E_B^m|\beta_m\rangle, \langle\langle\beta_m|H_B = \langle\langle\beta_m|E_B^m, \tag{2}$$

where $|\alpha_m\rangle$ ($|\beta_m\rangle$) and $\langle\langle\alpha_m|$ ($\langle\langle\beta_m|$) are the right and left eigenvectors of $H_A$ ($H_B$) with eigenvalue $E_A^m$ ($E_B^m$). The right eigenvectors of the total system $\mathcal{H}$ satisfy $\mathcal{H}|\psi_n\rangle = E_n|\psi_n\rangle$. We denote the spectra of $H_A$ and $H_B$ as $\eta(H_A)$ and $\eta(H_B)$, respectively. It is straightforward to see that $\eta(\mathcal{H}) = \eta(H_A) \cup \eta(H_B)$ and $\kappa$ plays no role in $\eta(\mathcal{H})$. When $\eta(H_A) = \eta(H_B)$, the system satisfies the ED condition. (Rigorously, the ED condition also requires $S_{nn} = \langle\langle\alpha_n|\kappa|\beta_n\rangle \neq 0$, which generically holds.) Under the ED condition, $\eta(\mathcal{H})$ entirely consists of order-2 EPs and $|\psi_n\rangle$ collapses to that of subsystem $A$ and has no distribution in system $B$, i.e., $|\psi_n\rangle = \begin{pmatrix} |\alpha_n\rangle \\ 0 \end{pmatrix}$. In other words, each state in $H_A$ forms an EP with a state from $H_B$, so the Hilbert space of $\mathcal{H}$ is $N$-fold defective and loses half the span. Note that this condition is distinct from the well-known order-$(N+1)$ EP: although it is also characterized by $N$-fold defectiveness, a higher-order EP is a single degenerate point with one energy, whereas the ED spectral coverage is the same as $\eta(H_A)$. There is no fundamental constraint on the number of bands therein and the bandwidths thereof.

## 3 | Frequency-domain EDGF

### 3.1 General properties

As a high-dimensional generalization of EP, ED has properties unknown to both Hermitian and conventional non-Hermitian systems. We first systematically study the steady-state response under the ED condition. The general definition of the frequency-domain GF is $\mathcal{G}(\omega) = (\omega - \mathcal{H})^{-1}$, where $\omega \in \mathbb{R}$ is the excitation (angular) frequency. Inserting the Hamiltonian Eq. (1), we have

$$\mathcal{G}(\omega) = \begin{pmatrix} G_A & G_{AB} \\ 0 & G_B \end{pmatrix}, \tag{3}$$

where $G_A = (\omega - H_A)^{-1}$, $G_B = (\omega - H_B)^{-1}$, and $G_{AB} \coloneqq G_A \kappa G_B$. From Eq. (3), it is immediately clear that a source at subsystem A generates a response in A given by $G_A$, but nothing in B. However, a source at subsystem B produces responses in both A and B, given by $G_{AB}$ and $G_B$, respectively. To further analyze the responses, we express the GFs using a Lehmann spectral expansion of $\mathcal{G}(\omega)$, in which both eigenvectors and Jordan vectors are required under the ED condition [28, 29]. However, for our unidirectionally coupled model, we can perform the spectral expansion of the GF using the eigenvectors of the subsystems A and B (subsystems A and B do not have EPs). The diagonal blocks are

$$G_A(\omega) = \sum_m \frac{|\alpha_m\rangle\langle\langle\alpha_m|}{\omega - E_A^m} \text{ and } G_B(\omega) = \sum_m \frac{|\beta_m\rangle\langle\langle\beta_m|}{\omega - E_B^m} \tag{4}$$

and the super-diagonal block is

$$G_{AB}(\omega) \coloneqq G_A \kappa G_B = \sum_m \sum_{m'} \frac{|\alpha_m\rangle\langle\langle\beta_{m'}| S_{mm'}}{(\omega - E_A^m)(\omega - E_B^{m'})}, \tag{5}$$

where $S_{mm'} = \langle\langle\alpha_m|\kappa|\beta_{m'}\rangle$. We further invoke the conditions of ED: $\eta(\mathcal{H}) = \eta(H_A) = \eta(H_B)$, or $E_A^m = E_B^m = E_m$, which lead to

$$G_{AB}(\omega) = \sum_m \frac{|\alpha_m\rangle\langle\langle\beta_m| S_{mm}}{(\omega - E_m)^2} + \sum_{m \neq m'} \frac{|\alpha_m\rangle\langle\langle\beta_{m'}| S_{mm'}}{(\omega - E_A^m)(\omega - E_B^{m'})}. \tag{6}$$

Equation (6) underpins the unique properties of the EDGF: its super-diagonal block consists of the spectral sum of second-order super-Lorentzian line shapes with a quadratic dependence on $\omega^{-1}$ and conventional Lorentzian line shapes dependent on $\omega^{-1}$. Compared to standard Lorentzian line shapes, there are two significant differences. First, the response magnitudes are significantly greater near resonance. Second, the response has no sign change across a resonance, and the response phase jumps by $2\pi$ instead of $\pi$.

To better understand the influence of the EDGF's poles on its amplitude and phase, we introduce a global loss $-i\gamma$ to the system, such that the eigenenergy becomes $E_m = \mathrm{Re}(E_m) - i\gamma$, which lies below the real axis. Under small loss $\gamma \to 0$, a source on subsystem A and B respectively generates responses that follow $\lim_{\gamma\to 0}\left(\frac{|\alpha_m\rangle}{\gamma}, 0\right)^{\mathrm{T}}$ and $\lim_{\gamma\to 0}\left(\frac{s_{mm}|\alpha_m\rangle}{\gamma^2}, \frac{|\beta_m\rangle}{\gamma}\right)^{\mathrm{T}}$. Therefore, regardless of where the source is located, the system's response has a dominant distribution on subsystem A for a small loss. The phase jump of the EDGF is related to the order of its pole; when the excitation frequency sweeps across a pole of order $p$, the phase of the EDGF changes by $p\pi$. This is because, for a pole in $(\omega - E_m)^{-p}$, its phase is $p\varphi_m(\omega)$, where the angle $\varphi_m(\omega)$ is illustrated in Fig. 1 (a). When the excitation frequency sweeps across $E_m$, the change in $\varphi_m(\omega)$ is approximately $\pi$, so the change in the phase of $(\omega - E_m)^{-p}$ is approximated as $p\pi$. Likewise, when passing through a $z$-th order zero, the phase also changes by $z\pi$.

The anomalous quadratic GF was theoretically reported in previous works on an individual EP that produces a single super-Lorentzian response peak [28, 29]. In the case of ED, the quadratic GF can appear for the first time in a spectral continuum, leading to broadband, unconventional steady-state responses. To clearly demonstrate its consequences, we compare two simple systems. The first one is a simple Hermitian single atomic chain with hopping $\tau$ under an open boundary condition (OBC). Its Hamiltonian reads $H_1 = \sum_{n=1}^{N-1} \tau\left(a_{n+1}^\dagger a_n + \mathrm{h.c.}\right) - \sum_{n=1}^{N} i\gamma a_n^\dagger a_n$, where $-i\gamma$ is the global dissipative rate, and $a_n^\dagger, a_n$ are the creation and annihilation operators at site $n$ [Fig. 1 (c)]. The second system is a double chain where both the upper (A) and lower (B) chains are identical Hermitian single atomic chains with hopping $\tau$, and the two chains are unidirectionally coupled from chain B to chain A via $u$ [Fig. 1 (d)]. The Hamiltonian reads $\mathcal{H}_2 = \begin{pmatrix} H_1 & uI \\ 0 & H_1 \end{pmatrix}$, where $I$ is an identity matrix. The single and double-chain systems apparently have the same spectrum. Because $H_2$ has a block-triangular form, it decomposes into $N$ different copies of $2 \times 2$ Jordan cells. As such, $H_2$ satisfies the condition of ED, with its entire spectrum formed by order-2 EPs with different energies. In summary, the key differences between $H_1$ and $\mathcal{H}_2$ are: $H_1$ has a conventional Hermitian spectrum and a set of $N$ orthonormal eigenstates. $\mathcal{H}_2$ has a non-Hermitian ED spectrum, is $N$-fold defective, but the non-defective part of its Hilbert space is spanned by $N$ non-defective orthonormal eigenstates. As such, comparison of these two systems provides a clear insight into the unique response properties under the ED condition.

The semi-continuous responses of these two systems, each with 20 unit cells, are computed using the GFs. The response of the single chain is shown in Fig. 1(d), where both the source and probe are placed at the first site, and a global dissipation $\gamma = 0.01$ is present. Due to the small dissipation, each excitation frequency $\omega = \mathrm{Re}(E_m)$ can be approximately regarded as a pole of the GF, whose modulus exhibits a peak proportional to $1/\gamma$ and phase changes by approximately $\pi$. As a result, a condition called anti-resonance, underpinned by the zeros in the GF, appears between every resonant peak. At anti-resonance, the system's response to external excitation vanishes because the responses contributed by the nearby resonances are out of phase in the intermediate frequencies and completely cancel each other at the anti-resonant frequency. For the double-chain system, when both the probe and source are located on the same chain, the responses are given by two diagonal blocks of the EDGF [Eq. (3)], and they are identical to those of the single chain, as shown in Fig. 1 (e, f). When the source and probe are placed at chain B and A, respectively, the response is determined by the super-diagonal block of the EDGF. Due to the orthogonality and normalization of the eigenstates of the upper and lower chains, the first (second) summation in Eq. (6) is non-zero (vanishing), So the response over the full spectrum consists of quadratic, super-Lorentzian peaks. This result is shown in Fig. 1(g). Because all poles are second order, the response peaks follow a quadratic relation, i.e., $1/\gamma^2$, and hence are two orders of magnitude greater in height than standard peaks.

We further compare the ED's broadband response enhancement with a single-EP model. The single EP model is constructed by deploying a coupling profile $\kappa_{1\mathrm{EP}} = |\psi_m\rangle\langle\chi|$, where $|\psi_m\rangle$ is the m-th eigenmode of chain A and $|\chi\rangle = |\psi_m\rangle + |0\rangle$, where $|0\rangle$ is a vector that only has support at site-0. (Note that this system has a rather complicated inter-chain coupling profile and cannot be presented by Fig. 1(d). It may not be realistic for realization. We only consider it for the comparison purpose.) In this case, the system only has one EP2 formed by the pair of modes with eigenenergy $E_m$, and the rest are doubly degenerate without forming EP2s. As before, we excite at $B_0$ and detect the response at $A_0$. Now, the response is $G_{AB}(\omega; A_0, B_0) = \frac{\psi_m(0)}{\omega - E_m}\left[\sum_n \frac{\langle\chi|\psi_n\rangle\psi_n^*(0)}{\omega - E_n}\right]$. Therefore, we can immediately derive that $E_m$ is a second-order pole of GF, and the phase of the response changes by $2\pi$ after passing through the EP, while all other peaks are due to the first-order poles, as shown in Fig. 1(b).

The super-diagonal block of the EDGF has another prominent feature: the phase of the response changes by $2\pi$ across each resonance, which is also caused by the quadratic form. As such, the GF does not have zeros, and the response phase does not wind back between the peaks. Anti-resonance is consequently absent. Therefore, when $\omega$ sweeps through all the poles, the phase winding is $2N\pi$.

The quadratic response in the EDGF also makes it more susceptible to dissipation than standard GFs. The reason also lies in the quadratic term $1/\gamma^2$. With significant $\gamma$, e.g., at the order of unity, the response peaks are considerably suppressed in magnitude, and the phase changes are less pronounced and much smoother. This is because excitation with real-frequency signals is too far from the EDGF's poles. For the same reason, zeros in standard GF cannot be exactly reached, so anti-resonances are not seen. Consequently, the differences between standard GF and EDGF are less pronounced, as shown in Fig. 1(h-j). In other words, the unique properties of EDGF are easily observable under low-loss conditions.

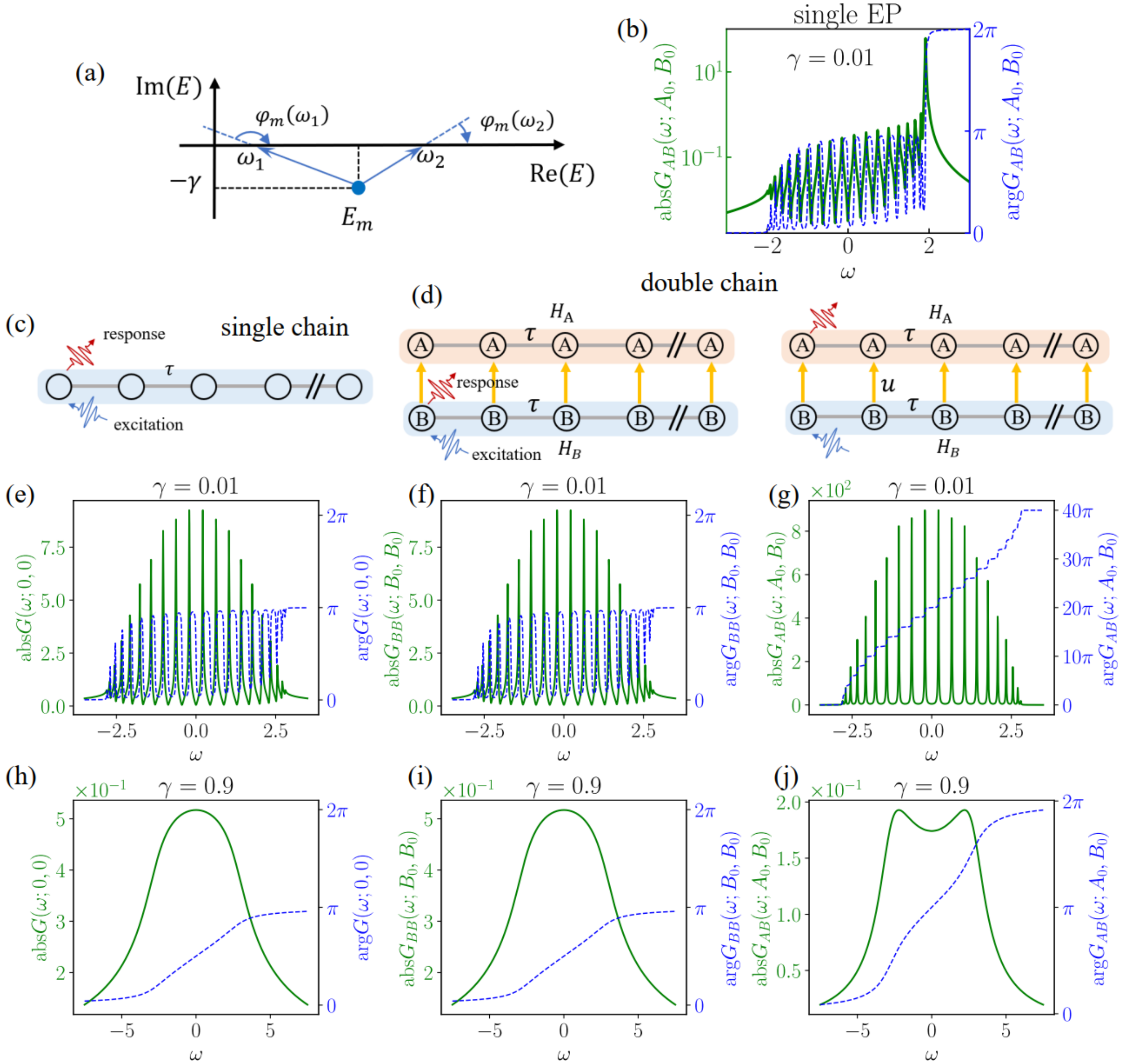

FIG. 1. (a) The relation of GF's phase and the excitation frequency. The blue dot represents a pole $E_m$ on the complex plane. (b) The frequency response of a double-chain lattice system with a single second-order EP, formed by the last pair of states in the band. (c) The Hermitian single atomic chain. (d) The ED-sustaining double chain model consists of two identical, unidirectionally coupled single atomic chains. The blue and red arrows indicate the positions of source and probe, respectively. (e) The frequency response for single chain. (f, g) The responses at chain B (f) and chain A (g) of the double-chain model. In (e-g), the global loss is $\gamma = 0.01$. (h-j) The responses of the single atomic chain (h) and the double chain (i ,j) for $\gamma = 0.9$. Other parameters are $\tau = 1$, $u = 1$ and $N = 20$.

### 3.2 Loss-revival of defective skin effect

The emergence of ED only requires the two subsystems to have an identical spectrum. In other words, the Hamiltonians $H_A$ and $H_B$ can be entirely different. This characteristic, combined with the properties of EDGF, can impart unexpected phenomena. Here, we report a surprising loss-revival of defective skin effect (LRDSE). The NHSE refers to the boundary accumulation of a macroscopic number of OBC eigenstates. It is commonly associated with a point gap in the complex PBC spectrum: the PBC bands form a loop winding around a reference energy, whereas the OBC spectrum generally collapses to a distinct curve or line segment [34, 35]. This spectral mismatch reflects the breakdown of conventional Bloch band theory and gives rise to skin-localized states. As a way to introduce non-Hermiticity, loss has been playing a central role in NHSE, e.g., inducing complex gauge fields [36]. Previous studies have shown that loss can impact the skin-mode responses by affecting the positions of poles in the GF [37, 38]. The LRDSE reported here is distinct from these known phenomena: it describes the re-emergence of defective skin modes in the steady-state response under the ED condition.

Consider an ED-sustaining system consisting of a finite-size Hermitian lattice and a finite-size non-Hermitian lattice with NHSE, as shown in Fig. 2(a). Chain B is a Hatano-Nelson chain with leftward and rightward hopping $t_L$ and $t_R$ with $t_L \neq t_R$. Its Hamiltonian reads $H_B = \sum_{n=1}^{N-1}\left(t_R a_{n+1}^\dagger a_n + t_L a_n^\dagger a_{n+1}\right) - \sum_{n=1}^{N} i\gamma a_n^\dagger a_n$. The Bloch band corresponding to this model has a point gap with non-trivial winding. Consequently, the model is affected by NHSE and all its eigenmodes are edge-localized skin modes. Chain A remains the same Hermitian single atomic chain with reciprocal hopping $\tau = \sqrt{t_L t_R}$, such that the OBC spectra of the two chains are identical. The two chains are one-way coupled from B to A via $u$. The PBC spectrum consists of a point gap (the green loop in Fig. 2 (b-e), originating from chain $B$) and a line segment overlapping with the OBC spectrum (originating from chain $A$). Clearly, only the green loop contributes to spectral winding. Therefore, all OBC energies lying inside the point gap have nonzero winding numbers. However, because the system is under ED, the eigenmodes are $|\psi_n\rangle = (|\alpha_n\rangle \quad 0)^{\mathrm{T}}$, *viz.* only the eigenmodes of chain A, which are extended states, are non-defective. All the skin modes associated with chain B are defective.

The properties of the eigenstates give the impression that extended states dominate the system's responses. However, the reality is far richer, thanks to the quadratic part of the EDGF. Place a point source at site $B_0$. The responses at coordinate $x$ on chains A and B are $G_{AB}(\omega; x, x') = G_{AB}(\omega; A_x, B_0)$ and $G_B(\omega; x, x') = G_B(\omega; B_x, B_0)$, respectively. Here, $x$ and $x'$ denote the position of the detector and the source, respectively. And $A_x$, $B_x$ represent the site at coordinate $x$ on chain A and B, respectively. According to the EDGF [Eq. (3-6)], when the loss $\gamma$ is small, $\omega \in \mathrm{Re}(E_n)$ almost coincides with the OBC spectrum, as shown in Fig. 2(b). And the responses on chain A and B scale as $S_{nn}|\alpha_n\rangle/\gamma^2$ and $|\beta_n\rangle/\gamma$, respectively. Therefore, the response is extended (right-edge localized) for chain A (B) [Fig. 2(f)]. It is worth noting that, since the eigenstates corresponding to different eigenvalues of the upper and lower chains are no longer orthogonal, both terms in Eq. (6) contribute to the response $G_{AB}$. However, when the loss is small, the response of the system is dominated by the first summation consisted of second-order super-Lorentzian line shape, characterized by enhanced magnitude, and sequential $2\pi$ phase shifts across each resonance. The response on chain A averaged over all excitation frequencies within the PBC loop, i.e., $\langle |G_{AB}| \rangle_{\mathrm{in}} = \frac{1}{N_\omega} \sum_{\omega_{\mathrm{in}}} G_{AB}(\omega_{\mathrm{in}}; A_x, B_0)$, is shown in Fig. 2(j), where $\omega_{\mathrm{in}}$ represents a frequency inside the PBC loop, and $N_\omega$ represents the number of frequencies inside the loop. However, when $\gamma$ increases, a phenomenon that we call LRDSE appears: when $\omega$ lies within the PBC point gap [Fig. 2(c-d)], the responses in both chains have an identical profile [Fig. 2(g-h)]: a typical skin-effect profile manifests with greater response magnitudes towards the right edge. This constitutes a highly counterintuitive phenomenon. The defects of skin modes unique to ED are not affected by the increasing loss. However, loss can induce a skin-type response that does not exist in the system's eigenstates, i.e., loss can "revive" the defective skin effect. To understand the LRDSE, we note that, due to the purely one-way interchain hopping from chain B to A, chain A is essentially driven by the response profile of chain B. As such, we observe that the response of chain A within the region marked by the magenta lines transitions from loss-induced decaying to a skin effect towards to right edge, as shown in Fig. 2(g-h). The transition frequency is exactly where the PBC spectral loop intersects with the real-energy axis, i.e., $\omega = \pm(t_R + t_L)\sqrt{1 - \gamma^2/(t_R - t_L)^2}$, which was predicted in ref. [25]. Figure 2(k-l) shows the average frequency response inside and outside the loop. The average frequency response outside the loop is defined as $\langle |G_{AB}| \rangle_{\mathrm{out}} = \frac{1}{N_\omega} \sum_{\omega_{\mathrm{out}}} G_{AB}(\omega_{\mathrm{out}}; A_x, B_0)$, where $\omega_{\mathrm{out}}$ represents a frequency outside the PBC loop, and $N_\omega$ represents the number of frequencies outside the loop. Clearly, the response inside the loop exhibits a skin effect, while the response outside the loop is localized at the excitation position. Further increase of $\gamma$ causes the PBC spectral loop to drop entirely below the real-energy axis, i.e., $\gamma \geq |t_R - t_L|$ [Fig.

2(e)], and NHSE can no longer be observed in the response of chain B. As such, it also disappears for chain A, as shown in Fig. 2(i, m).

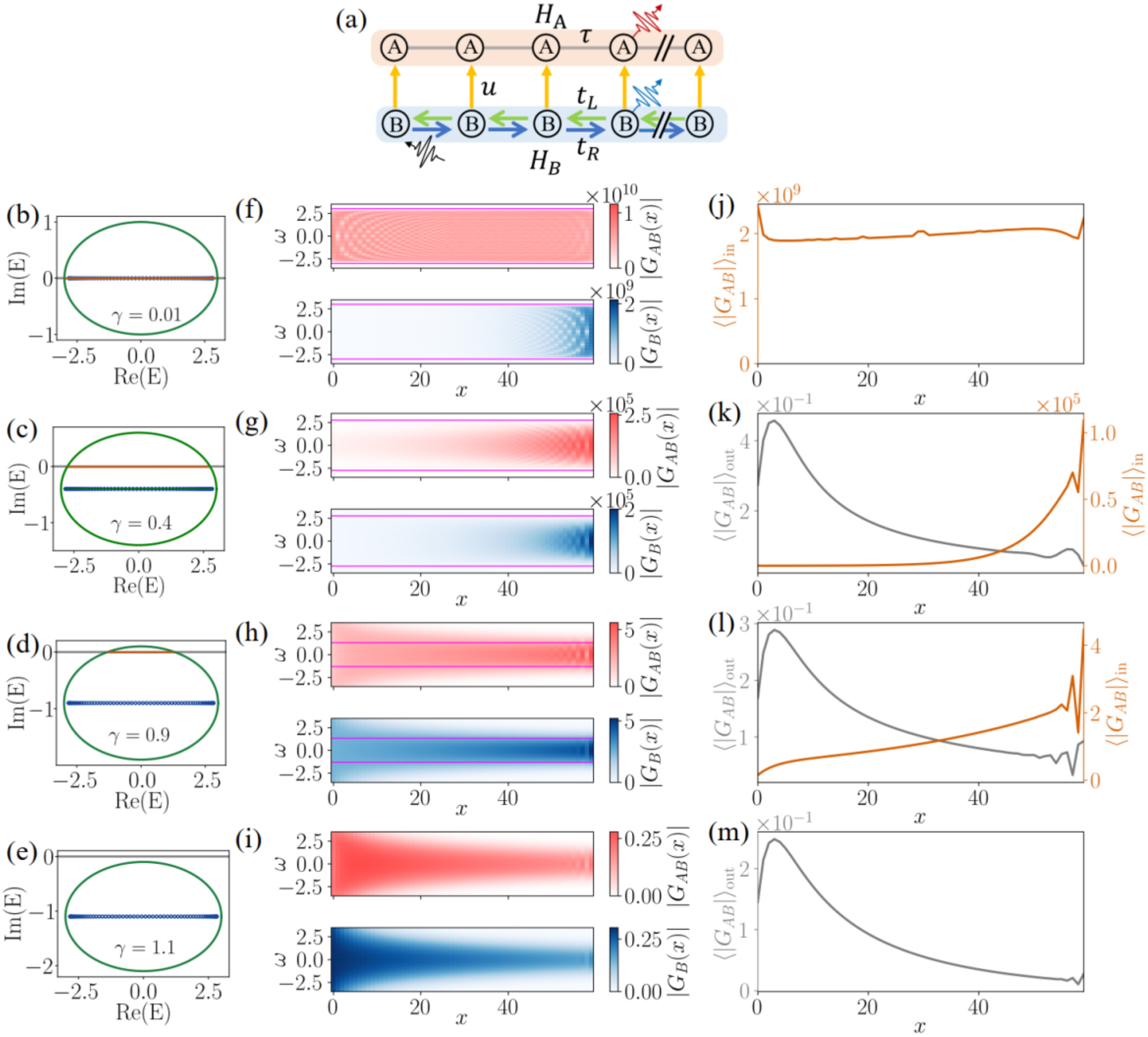

FIG. 2. (a) The double chain models where chain A is a Hermitian single atom chain and chain B is a non-Hermitian Hatano-Nelson chain. The two chains are unidirectionally coupled by $u$. (b-e) the PBC (green line) and OBC (blue dots) spectrum of Hatano-Nelson model under different global losses. (f-i) The magnitudes of the responses at $A_x$ (red) and $B_x$ (blue) with the source at $B_0$ for different losses as indicated in (b-e). The LRDSE is seen in (g) and (h). (j-m) The averaged response on chain A over excitation frequencies inside and outside the PBC loop for different losses, where the grey and orange curves correspond to source frequency outside and inside the PBC loop, respectively. Other parameters are $\tau = \sqrt{2}$, $t_L = 1$, $t_R = 2$ and $u = 1$.

**4 | Time-domain EDGF**

### 4.1 General properties

As discussed, the frequency-domain EDGF is distinct from conventional GF, the same can be expected for the EDGF in the time domain. We perform Fourier transform on $\mathcal{G}(\omega)$ to obtain the time-domain EDGF, i.e., $\mathcal{G}(t) = \oint_C \frac{d\omega}{2\pi} \mathcal{G}(\omega) e^{-i\omega t}$, where the contour $C$ encloses the complex energy spectrum. Just like Eq. (3), time-domain EDGF $G(t)$ also takes a block-triangular form

$$\mathcal{G}(t) = \begin{pmatrix} G_A(t) & G_{AB}(t) \\ 0 & G_B(t) \end{pmatrix}. \tag{7}$$

The diagonal blocks are $G_A(t) = -\sum_m |\alpha_m\rangle\langle\langle\alpha_m| e^{-iE_A^m t}$, $G_B(t) = -\sum_m |\beta_m\rangle\langle\langle\beta_m| e^{-iE_B^m t}$. These two terms describe ordinary time-harmonic response. The super-diagonal block has more interesting features

$$G_{AB}(t) = \sum_{E_A^m \neq E_B^{m'}} |\alpha_m\rangle\langle\langle\beta_{m'}| S_{mm'} e^{-i\tilde{E}t} \frac{\sin(\Delta t)}{\Delta} - \sum_{E_A^m = E_B^{m'}} |\alpha_m\rangle\langle\langle\beta_{m'}| S_{mm'} t e^{-iE_m t}. \tag{8}$$

where $\Delta = (E_A^m - E_B^{m'})/2$ and $\tilde{E} = \left(E_A^m + E_B^{m'}\right)/2$. The first term in Eq. (8) describes a beating in the response amplitude, which is clearly caused by the difference in the eigenvalues of two chains. What is interesting lies in the second term, which persists under the condition of ED, i.e., $E_A^m = E_B^m = E_m$ and $S_{mm} \neq 0$. The second term has a linear dependence on time $t$, in addition to the expected time-harmonic oscillation. Clearly, this leads to a continuous, linear-in-time amplification for the wavefunction. We remark that the time-linear amplification effect is a purely ED-induced phenomenon, and it does not relate to the imaginary parts of the complex eigenenergies. Recently, a similar time amplification behavior was also found in a non-Hermitian square-root topological insulator satisfying ED [39].

### 4.2 Linear temporal amplification effect

To further investigate the time amplification of EDGF, we define $F(t) = \frac{1}{N} \mathrm{Tr}(G_{AB}^\dagger G_{AB})$ to characterize the strength of the Green's function at time $t$, which captures the energy of the wave across the entire space. A global dissipation $-i\gamma$ is introduced. $F(t)$ first increases and then decays, reaching its maximum value at $t = 1/\gamma$, which scales as $\frac{M}{N\gamma^2 e^2}$. Here, $M$ is the number of EPs and $N$ is the size of the matrix $G_{AB}$. In our ED-sustaining system, $M \sim \mathcal{O}(N)$, which results in a significant time amplification effect. In comparison, the time amplification in non-Hermitian systems with individual EP cannot be easily observable because $M \sim \mathcal{O}(1)$.

To better demonstrate the time-linear amplification effect, we consider the same model as shown in Fig. 1(c), i.e., double chain system with two Hermitian single-atom lattices with unidirectional interchain coupling [Fig. 3(b)]. We set a global dissipation to $\gamma = 0.2$. We prepare the initial state by injecting a Gaussian wavepacket at the center of chain B, i.e., $\varphi_B(x,0) = Ze^{-\sigma^2(x-x_c)^2/2}e^{ik_c(x-x_c)}$, where $Z = \pi^{-1/4}\sigma^{1/2}$ is the normalization factor, $k_c$ and $x_c$ are the center positions of the wavepacket in the momentum and real space, respectively, and $\sigma$ is the width of the wavepacket in momentum space. Through straightforward calculation, we can obtain the distribution of chain B is given by $|\varphi_B(x,t)|^2 \propto \exp\left(-2\gamma t - \frac{(x-x_c-v_g t)^2\sigma^2}{1+d^2t^2\sigma^4}\right)$, and the distribution on chain A is $|\varphi_A(x,t)|^2 \propto t^2|\varphi_B(x,t)|^2$, where $v_g = \frac{\partial E_k}{\partial k}|_{k=k_c} = -2\tau\sin k_c$, with $E_k = 2\tau\cos k - i\gamma$ being the Bloch eigenenergy, is the group velocity of the wavepacket, and $d = \frac{\partial^2 E_k}{\partial k^2}|_{k=k_c} = -2\tau\cos k_c$. Here, owing to theorthogonality of the eigenstates with different eigenvalues from the upper and lower chains, only the second term in Eq. (8) contributes to the evolution. The evolutions of wavepackets on chain B and chain A are shown in Fig. 3(d) and (e), respectively. Here, the initial momentum of the wavepacket is $k_c = 0$ so the initial group velocity is $v_g = 0$. It is seen that the evolution of the wavepacket on chain B is identical to that of an ordinary Hermitian chain [Fig. 3(c)], and it disregards the existence of chain A. In contrast, the evolution on chain A has an additional amplification factor proportional to $t^2$, which originates from the contributions of ED. We compute the total energy of the wave in the two chains as $P_{A,B} = \int dx\left|\varphi_{A,B}(x,t)\right|^2$. The results are $P_B = e^{-2\gamma t}$ and $P_A = t^2e^{-2\gamma t}$, respectively plotted in Fig. 3 (g) and (h). Obviously, the evolution of the total energy in chain B is identical to that of a single Hermitian chain [Fig. 3(f)]. The amplification effect is clearly observed in chain A but not in chain B, where $P_A$ exhibits an amplification that scales as $t^2$ in the short-time limit, as shown in the insert of Fig. 3(h), it then reaches its maximum value $\gamma^{-2}e^{-2}$ at $t = 1/\gamma$ due to the competition between ED and dissipation, and eventually overpowered by the dissipation in the long-time limit. Figure 3 (h) further compares the wavepacket energy of the single EP system and the ED system mentioned in Sec **3.1**. $P_A(t)$ of the single EP system is shown as the red dashed line in Fig. 3(h). Both models contain the same local EP2 factor, $t^2e^{-2\gamma t}$, but the ED system exhibits a much stronger total amplification because the entire spectral content of the wavepacket contributes, whereas the isolated EP2 amplifies only one modal component. This comparison demonstrates that ED enables wavepacket dynamics and response phenomena involving an extended band of coalesced modes, which cannot be accessed using a single isolated EP2.

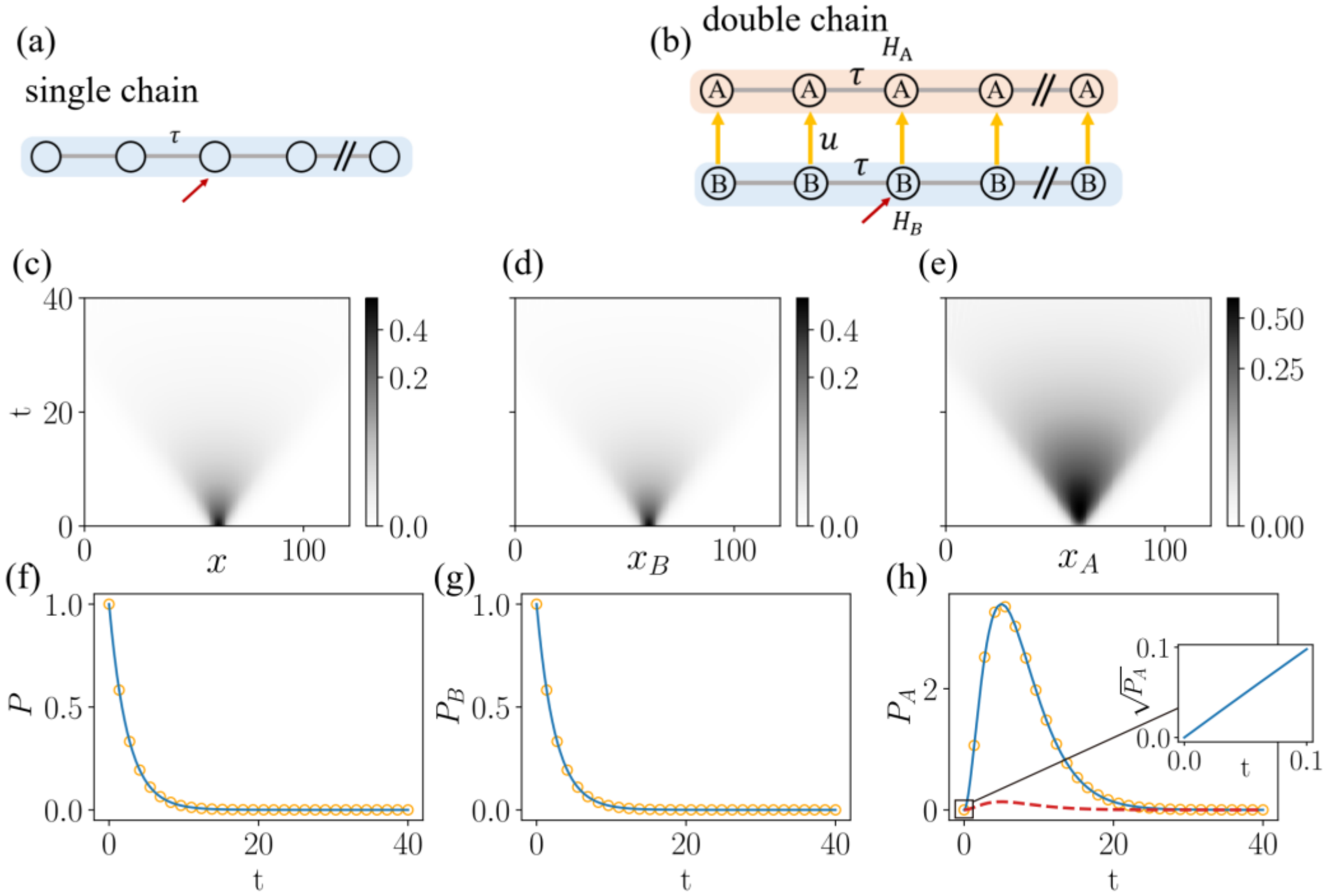

FIG. 3. The schematic diagrams of a Hermitian single-atom chain (a), and double-chain model consists of two identical, unidirectionally coupled single-atom chains (b). (c-e) The evolution of the wavepacket for single chain (c) and chain B (d) [chain A (e)] of double chain when the center of the initial wave packet is at the center of the reciprocal single-atom chain and chain B, respectively. We set the center positions of the initial wavepacket in the momentum is $k_c = 0$, i.e., the group velocity is zero. (f-h) The evolution of the wave packet energy $P$ for single chain (f) and $P_B$ (g) [$P_A$ (h)] for the double chain. The insert in (h) shows the short-time evolution of $\sqrt{P_A}$. The blue lines in (f-h) represent the numerical results, while the orange hollow circles represent the theoretical results. The red dashed line in (h) represents the wave packet energy evolution of the single EP system (same as the one shown in Fig. 1(b)). Other parameters are $\tau = 1,\ u = 1$, and the lengths of single chain and double chains are 122.

We also consider the time-domain EDGF in the presence of NHSE. Here, both chains are identical Hatano-Nelson models, as depicted in Fig. 4(b). Here, we apply a large global dissipation of $\gamma = 0.9$. We will observe that under this loss, the response of chain B has been suppressed, but chain A can still exhibit gain due to the linear time factor. The initial state is a Gaussian wave packet injected in the middle of chain B. The initial group velocity of the wavepacket is zero, so the displacement of the wavepacket center is entirely caused by the NHSE. The evolutions of wavepackets on chain B and chain A are shown in Fig. 4(d), (e), respectively. Due to the biorthogonality of the left and right eigenvectors for the non-Hermitian case, $S_{mm'}$ are zero for all $m \neq m'$, and therefore only the second term in Eq. (8) contributes to the evolution. As expected, the evolution of the wavepacket on chain B is dominated by the rightward NHSE and is identical to that of the ordinary Hatano-Nelson chain [Fig. 4(c)]. The wavepacket evolution in chain B exhibits exponential decay, which is clearly caused by the global dissipation. Upon reaching the edge, the wavepacket experiences amplification by the NHSE (the hump in Fig. 4(g) near $t = 20$), then eventually vanishes. These results are identical to those of a single Hatano-Nelson chain, which are shown in Fig. 4(f) as reference. In contrast, the evolution on chain A exhibits a significant amplification [Fig. 4(e)], which originates from the time-linear amplification in the EDGF. It reaches over 10 folds of the injected energy, before eventually taken over by loss [Fig. 4 (h)], thereby allowing us to clearly observe the skin effect that is otherwise suppressed by large dissipation.

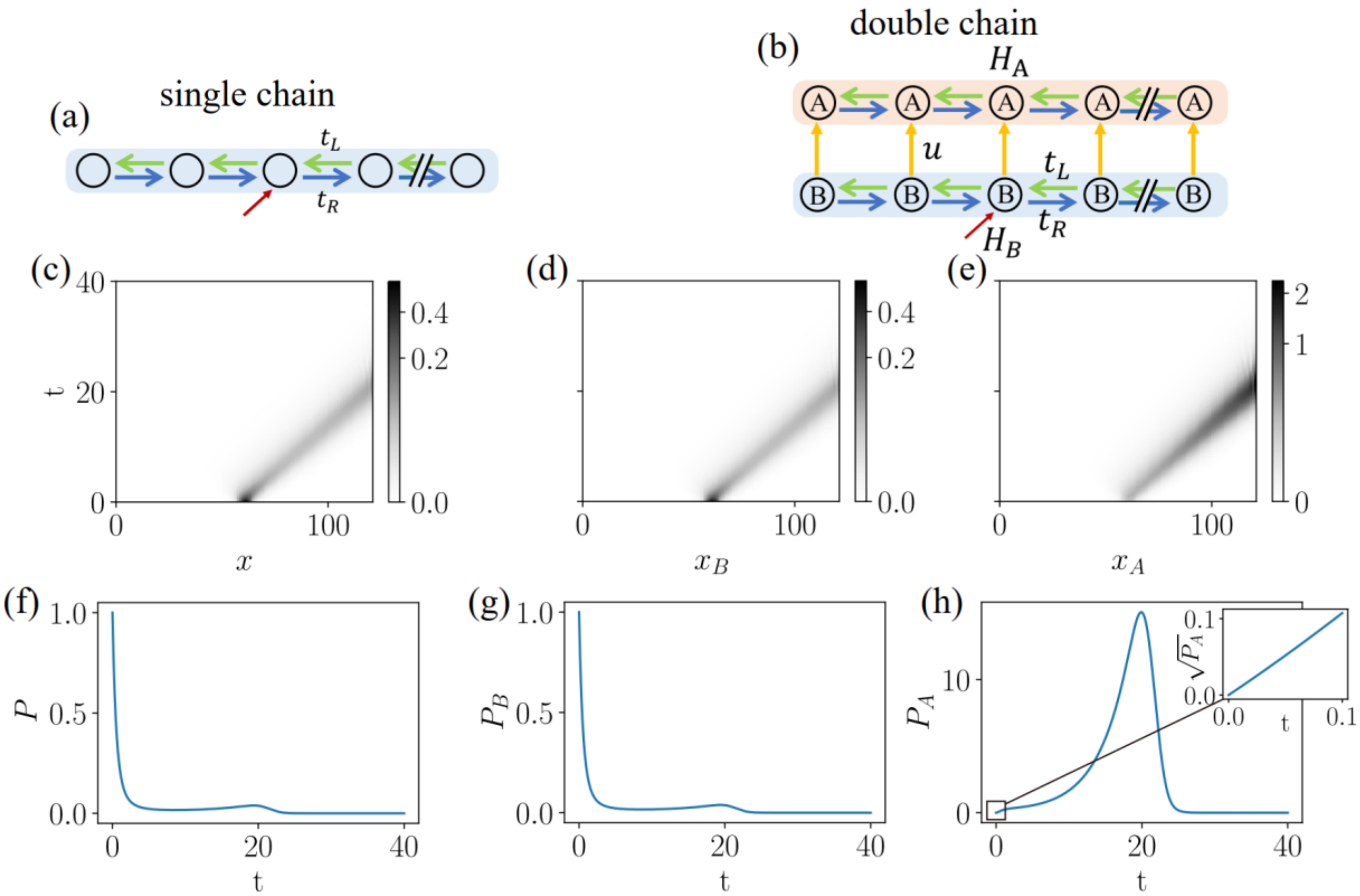

FIG. 4. The schematic diagrams of a single Hatano-Nelson chain (a), and two unidirectionally coupled Hatano-Nelson chains (b). (c) The evolution of the wavepacket for the single Hatano-Nelson chain. (d, e) The evolution of the wavepacket in chain B (d) and chain A (e) of double chain. In (c-e), the initial wave packet is at the center of the single Hatano-Nelson chain and chain B of double chain, respectively. We set the center positions of the initial wavepacket in the momentum is $k_c = 0$. (f-h) The evolution of the wave packet energy $P$ (f) for single chain and $P_B$ (g) [$P_A$ (h)] for double chain. The insert in (h) shows the short-time evolution of $\sqrt{P_A}$. Other parameters are $u = 1$, $t_L = 1$, $t_R = 2$, and the lengths of single chain and double chain are 122.

**5 | Conclusion**

In this work, we have studied the steady-state and time-domain responses of non-Hermitian systems at ED by analysing the EDGFs. The properties of the EDGFs reveal several untapped potentials in novel non-Hermitian phenomena enabled by ED. The frequency-domain EDGF shows that the responses are dominated by the spectral sum of super-Lorentzian line shapes with second-order poles, opening considerable new application potential. Perhaps the most obvious benefit comes from the peak magnitudes of the super-Lorentzian line shapes, which can benefit applications requiring strong intensity. We remark that because the ED is condition that can appear for a spectral continuum instead of a single resonance, the boost in response magnitudes can have a broad spectral coverage. The super-Lorentzian EDGF also features $2\pi$ phase changes across each resonance, whereas in standard GF, such phase change is $\pi$. Because the phase changes are crucial for determining the behaviours of resonances, such a difference will play significant role in future attempts in novel non-Hermitian phenomena. In particular, we anticipate new twists in controlling anti-resonances and Fano resonances, which underpin a large variety of phenomena and applications in acoustic and vibrational systems [40], metamaterials [41, 42], photonics [43, 44], plasmonics [45], nanotechnologies [46], and so on. For example, imagine building a metasurface based on ED. This metasurface can, in principle, impose phase manipulation that far exceeds $2\pi$, which can significantly expand the range of functionalities. For example, instead of wave steering, we may be able to substantially "slow" the propagation of a wavepacket using a very thin structure by imposing a large phase delay.

The EDGF also unveils new possibility in controlling NHSEs with loss, as demonstrated by the LRDSE. In the time-domain EDGF, we have identified a time-linear amplification effect, which may be leveraged to compensate dissipation or to enhance non-linear responses. The linear increase of the probability amplitude in the short-time limit is in fact a universal feature of quantum transition amplitudes within the energy-conserving approximation. Examples include spontaneous emission from an excited atom and elastic scattering in a weak field [47]. In the short-time regime the constituent waves have not yet de-phased, so the amplitude adds coherently and grows linearly with time, whereas at long times it transitions to exponential decay. Our ED system exhibits the same dynamical behavior, hinting a potential physical connection with these phenomena in quantum transition.

Many experimental platforms can be used to realize ED and investigate EDGF. The key requirement being the realization of unidirectional coupling. An obvious choice is the active mechanical lattice, which has been used to demonstrate many non-Hermitian phenomena [48-54], including recently the ED [30]. In non-Hermitian electrical circuits [55, 56], the two subsystems may be constructed from parallel arrays of LC resonators, with the unidirectional coupling realized using voltage followers and coupling capacitors. Because the voltage follower transmits the voltage signal forward while suppressing electrical feedback to the input node, it naturally realizes the required one-way interchain coupling. Unidirectional coupling can also be achieved in acoustic resonator systems using active microphone–amplifier–loudspeaker feedback[57], or in photonic and microwave systems using chiral waveguide-mediated interactions arising from spin–momentum locking [58]. These schemes provide practical routes for realizing the block-triangular structure underlying the exceptional-deficiency physics studied here. On the other hand, future efforts will be devoted to finding ED in non-Hermitian Hamiltonians without the block-triangular structure.

**Acknowledgement**

C. L. thanks Xulong Wang for helpful discussions. This work was supported by the National Natural Science Foundation of China (T2525002), the National Key R&D Program (2022YFA1404400), the Hong Kong Research Grants Council (RFS2223-2S01, 12301822, 12300925), and the Hong Kong Baptist University (RC-RSRG/23-24/SCI/01, RC-SFCRG/23-24/R2/SCI/12).

**Data Availability Statement**

The data that support the findings of this study are available from the corresponding author upon reasonable request.

## References

[1] C.M. Bender, Making sense of non-Hermitian Hamiltonians, Rep. Prog. Phys., 70 (2007) 947.
[2] Y. Ashida, Z. Gong, M. Ueda, Non-Hermitian physics, Advances in Physics, 69 (2020) 249–435.
[3] E.J. Bergholtz, J.C. Budich, F.K. Kunst, Exceptional topology of non-Hermitian systems, Rev. Mod. Phys., 93 (2021) 015005.
[4] K. Ding, C. Fang, G. Ma, Non-Hermitian topology and exceptional-point geometries, Nat. Rev. Phys., 4 (2022) 745–760.
[5] L. Feng, R. El-Ganainy, L. Ge, Non-Hermitian photonics based on parity–time symmetry, Nature Photonics, 11 (2017) 752–762.
[6] M.-A. Miri, A. Alù, Exceptional points in optics and photonics, Science, 363 (2019) eaar7709.
[7] L. Huang, S. Huang, C. Shen, S. Yves, A.S. Pilipchuk, X. Ni, S. Kim, Y.K. Chiang, D.A. Powell, J. Zhu, Acoustic resonances in non-Hermitian open systems, Nat. Rev. Phys., 6 (2024) 11–27.
[8] L. Feng, Z.J. Wong, R.-M. Ma, Y. Wang, X. Zhang, Single-mode laser by parity-time symmetry breaking, Science, 346 (2014) 972–975.
[9] B. Peng, K. Ozdemir, S. Rotter, H. Yilmaz, M. Liertzer, F. Monifi, C.M. Bender, F. Nori, L. Yang, Loss-induced suppression and revival of lasing, Science, 346 (2014) 328–332.
[10] A. Schumer, Y. Liu, J. Leshin, L. Ding, Y. Alahmadi, A. Hassan, H. Nasari, S. Rotter, D. Christodoulides, P. LiKamWa, Topological modes in a laser cavity through exceptional state transfer, Science, 375 (2022) 884–888.
[11] B. Peng, Ş.K. Özdemir, M. Liertzer, W. Chen, J. Kramer, H. Yılmaz, J. Wiersig, S. Rotter, L. Yang, Chiral modes and directional lasing at exceptional points, Proceedings of the National Academy of Sciences, 113 (2016) 6845–6850.
[12] M.P. Hokmabadi, A. Schumer, D.N. Christodoulides, M. Khajavikhan, Non-Hermitian ring laser gyroscopes with enhanced Sagnac sensitivity, Nature, 576 (2019) 70–74.
[13] J. Wiersig, Enhancing the sensitivity of frequency and energy splitting detection by using exceptional points: application to microcavity sensors for single-particle detection, Phys. Rev. Lett., 112 (2014) 203901.
[14] H. Hodaei, A.U. Hassan, S. Wittek, H. Garcia-Gracia, R. El-Ganainy, D.N. Christodoulides, M. Khajavikhan, Enhanced sensitivity at higher-order exceptional points, Nature, 548 (2017) 187–191.
[15] P.-Y. Chen, M. Sakhdari, M. Hajizadegan, Q. Cui, M.M.-C. Cheng, R. El-Ganainy, A. Alù, Generalized parity–time symmetry condition for enhanced sensor telemetry, Nat. Electron., 1 (2018) 297–304.
[16] Z. Dong, Z. Li, F. Yang, C.-W. Qiu, J.S. Ho, Sensitive readout of implantable microsensors using a wireless system locked to an exceptional point, Nat. Electron., 2 (2019) 335–342.
[17] Y.-H. Lai, Y.-K. Lu, M.-G. Suh, Z. Yuan, K. Vahala, Observation of the exceptional-point-enhanced Sagnac effect, Nature, 576 (2019) 65–69.
[18] W. Chen, Ş. Kaya Özdemir, G. Zhao, J. Wiersig, L. Yang, Exceptional points enhance sensing in an optical microcavity, Nature, 548 (2017) 192–196.
[19] J.-H. Park, A. Ndao, W. Cai, L. Hsu, A. Kodigala, T. Lepetit, Y.-H. Lo, B. Kanté, Symmetry-breaking-induced plasmonic exceptional points and nanoscale sensing, Nat. Phys., 16 (2020) 462–468.
[20] R. Kononchuk, J. Cai, F. Ellis, R. Thevamaran, T. Kottos, Exceptional-point-based accelerometers with enhanced signal-to-noise ratio, Nature, 607 (2022) 697–702.
[21] S. Assawaworrarit, X. Yu, S. Fan, Robust wireless power transfer using a nonlinear parity–time-symmetric circuit, Nature, 546 (2017) 387–390.
[22] S. Assawaworrarit, S. Fan, Robust and efficient wireless power transfer using a switch-mode implementation of a nonlinear parity–time symmetric circuit, Nature Electronics, 3 (2020) 273–279.
[23] J. Doppler, A.A. Mailybaev, J. Böhm, U. Kuhl, A. Girschik, F. Libisch, T.J. Milburn, P. Rabl, N. Moiseyev, S. Rotter, Dynamically encircling an exceptional point for asymmetric mode switching, Nature, 537 (2016) 76–79.

[24] H. Xu, D. Mason, L. Jiang, J. Harris, Topological energy transfer in an optomechanical system with exceptional points, Nature, 537 (2016) 80–83.
[25] W.-T. Xue, M.-R. Li, Y.-M. Hu, F. Song, Z. Wang, Simple formulas of directional amplification from non-Bloch band theory, Phys. Rev. B, 103 (2021) L241408.
[26] Z. Lin, H. Ramezani, T. Eichelkraut, T. Kottos, H. Cao, D.N. Christodoulides, Unidirectional invisibility induced by PT-symmetric periodic structures, Phys. Rev. Lett., 106 (2011) 213901.
[27] E. Rivet, A. Brandstötter, K.G. Makris, H. Lissek, S. Rotter, R. Fleury, Constant-pressure sound waves in non-Hermitian disordered media, Nat. Phys., 14 (2018) 942–947.
[28] J. Wiersig, Revisiting the hierarchical construction of higher-order exceptional points, Phys. Rev. A, 106 (2022) 063526.
[29] A. Hashemi, K. Busch, D. Christodoulides, S. Ozdemir, R. El-Ganainy, Linear response theory of open systems with exceptional points, Nat. Commun., 13 (2022) 3281.
[30] Z. Li, X. Wang, R. Cai, K. Shimomura, C. Lu, Z. Yang, M. Sato, G. Ma, Exceptional deficiency of non-Hermitian systems, Nature Physics, 22 (2026) 962–970.
[31] G.-F. Guo, X.-X. Bao, H.-J. Zhu, X.-M. Zhao, L. Zhuang, L. Tan, W.-M. Liu, Anomalous non-Hermitian skin effect: topological inequivalence of skin modes versus point gap, Communications Physics, 6 (2023) 363.
[32] D. Zhou, L. Nie, R. Hu, X. Wang, J. Wu, Z. He, K. Deng, Observation of anomalous non-Hermitian skin effect in electric circuits, Phys. Rev. B, 111 (2025) 224104.
[33] Q. Zhong, J. Kou, Ş. Özdemir, R. El-Ganainy, Hierarchical construction of higher-order exceptional points, Phys. Rev. Lett., 125 (2020) 203602.
[34] N. Okuma, M. Sato, Non-Hermitian Topological Phenomena: A Review, Annual Review of Condensed Matter Physics, 14 (2023) 83–107.
[35] R. Lin, T. Tai, L. Li, C.H. Lee, Topological Non-Hermitian skin effect, Frontiers of Physics, 18 (2023) 53605.
[36] Y. Li, C. Liang, C. Wang, C. Lu, Y.-C. Liu, Gain-Loss-Induced Hybrid Skin-Topological Effect, Phys. Rev. Lett., 128 (2022) 223903.
[37] Y.-M. Hu, Z. Wang, Green's functions of multiband non-Hermitian systems, Physical Review Research, 5 (2023) 043073.
[38] J. Huang, K. Ding, J. Hu, Z. Yang, Complex frequency fingerprint: Experimentally accessible method for detecting complex-valued eigenfrequencies, Physical Review B, 113 (2026) 075128.
[39] S. Bid, H. Schomerus, Exceptionally deficient topological square-root insulators, Physical Review Research, 8 (2026) L012031.
[40] F. Wahl, G. Schmidt, L. Forrai, On the significance of antiresonance frequencies in experimental structural analysis, Journal of Sound and Vibration, 219 (1999) 379–394.
[41] G. Ma, P. Sheng, Acoustic metamaterials: From local resonances to broad horizons, Science Advances, 2 (2016) e1501595.
[42] Y. Xu, Y. Fu, H. Chen, Planar gradient metamaterials, Nat. Rev. Mater., 1 (2016) 16067.
[43] M.F. Limonov, Fano resonance for applications, Adv. Opt. Photonics, 13 (2021) 703–771.
[44] M.F. Limonov, M.V. Rybin, A.N. Poddubny, Y.S. Kivshar, Fano resonances in photonics, Nature photonics, 11 (2017) 543–554.
[45] B. Luk'Yanchuk, N.I. Zheludev, S.A. Maier, N.J. Halas, P. Nordlander, H. Giessen, C.T. Chong, The Fano resonance in plasmonic nanostructures and metamaterials, Nat. Mater., 9 (2010) 707–715.
[46] A.E. Miroshnichenko, S. Flach, Y.S. Kivshar, Fano resonances in nanoscale structures, Reviews of Modern Physics, 82 (2010) 2257–2298.
[47] J.J. Sakurai, J. Napolitano, Modern quantum mechanics, Cambridge university press2020.
[48] W. Wang, X. Wang, G. Ma, Non-Hermitian morphing of topological modes, Nature, 608 (2022) 50–55.
[49] W. Wang, X. Wang, G. Ma, Extended State in a Localized Continuum, Physical Review Letters, 129 (2022) 264301.
[50] X. Cui, R.-Y. Zhang, X. Wang, W. Wang, G. Ma, C.T. Chan, Experimental Realization of Stable Exceptional Chains Protected by Non-Hermitian Latent Symmetries Unique to Mechanical Systems, Physical Review Letters, 131 (2023) 237201.

[51] W. Wang, X. Wang, G. Ma, Anderson Transition at Complex Energies in One-Dimensional Parity-Time-Symmetric Disordered Systems, Physical Review Letters, 134 (2025) 066301.
[52] X. Wang, G. Ma, Experimental measurement of non-Hermitian left eigenvectors, Frontiers of Physics, 20 (2025) 54202.
[53] C. Lu, X. Wang, G. Ma, Experimental Realization of Special-Unitary Operations in Classical Mechanics by Nonadiabatic Evolutions, Physical Review Letters, 135 (2025) 027201.
[54] M.-W. Li, J.-W. Liu, X. Wang, W.-J. Chen, G. Ma, J.-W. Dong, Topological Temporal Boundary States in a Non-Hermitian Spatial Crystal, Physical Review Letters, 135 (2025) 187101.
[55] T. Helbig, T. Hofmann, S. Imhof, M. Abdelghany, T. Kiessling, L.W. Molenkamp, C.H. Lee, A. Szameit, M. Greiter, R. Thomale, Generalized bulk–boundary correspondence in non-Hermitian topolectrical circuits, Nature Physics, 16 (2020) 747–750.
[56] C.-X. Guo, L. Su, Y. Wang, L. Li, J. Wang, X. Ruan, Y. Du, D. Zheng, S. Chen, H. Hu, Scale-tailored localization and its observation in non-Hermitian electrical circuits, Nature Communications, 15 (2024) 9120.
[57] Z. Chen, Z. Li, J. Weng, B. Liang, Y. Lu, J. Cheng, A. Alù, Sound non-reciprocity based on synthetic magnetism, Science Bulletin, 68 (2023) 2164–2169.
[58] S. Wang, B. Hou, W. Lu, Y. Chen, Z. Zhang, C.T. Chan, Arbitrary order exceptional point induced by photonic spin–orbit interaction in coupled resonators, Nat. Commun., 10 (2019) 832.